\documentclass[english,12pt]{article}

\usepackage{jheppub}

\usepackage[T1]{fontenc}
\usepackage[latin9]{inputenc}
\usepackage{amsfonts}
\usepackage{amsmath}
\usepackage{amssymb}
\usepackage{bbm}
\usepackage{cancel}
\usepackage{subcaption}
\usepackage{caption}
\usepackage{comment}
\usepackage{mathtools}
\usepackage{dsfont}
\usepackage{verbatim}
\usepackage{cancel}
\usepackage{blkarray}
\usepackage{graphicx}
\usepackage{slashed}
\usepackage{color}

\makeatletter

\usepackage{slashed}
\usepackage{physics}
\usepackage{tikz-cd}

\usepackage{hyperref}
\hypersetup{
	colorlinks,
	citecolor=blue,
	filecolor=blue,
	linkcolor=blue,
	urlcolor=blue,
	linktocpage=true
}

\newcommand\identity{1\kern-0.25em\text{l}}
\makeatother

\usepackage{booktabs}

\usepackage{wrapfig}

\usepackage{standalone}
\usepackage{tikz}

\tikzset{
pattern size/.store in=\mcSize, 
pattern size = 5pt,
pattern thickness/.store in=\mcThickness, 
pattern thickness = 0.3pt,
pattern radius/.store in=\mcRadius, 
pattern radius = 1pt}
\makeatletter
\pgfutil@ifundefined{pgf@pattern@name@_klt208kxf}{
\pgfdeclarepatternformonly[\mcThickness,\mcSize]{_klt208kxf}
{\pgfqpoint{0pt}{0pt}}
{\pgfpoint{\mcSize+\mcThickness}{\mcSize+\mcThickness}}
{\pgfpoint{\mcSize}{\mcSize}}
{
\pgfsetcolor{\tikz@pattern@color}
\pgfsetlinewidth{\mcThickness}
\pgfpathmoveto{\pgfqpoint{0pt}{0pt}}
\pgfpathlineto{\pgfpoint{\mcSize+\mcThickness}{\mcSize+\mcThickness}}
\pgfusepath{stroke}
}}
\makeatother

\tikzset{
pattern size/.store in=\mcSize, 
pattern size = 5pt,
pattern thickness/.store in=\mcThickness, 
pattern thickness = 0.3pt,
pattern radius/.store in=\mcRadius, 
pattern radius = 1pt}
\makeatletter
\pgfutil@ifundefined{pgf@pattern@name@_49wcz6le7}{
\pgfdeclarepatternformonly[\mcThickness,\mcSize]{_49wcz6le7}
{\pgfqpoint{0pt}{0pt}}
{\pgfpoint{\mcSize+\mcThickness}{\mcSize+\mcThickness}}
{\pgfpoint{\mcSize}{\mcSize}}
{
\pgfsetcolor{\tikz@pattern@color}
\pgfsetlinewidth{\mcThickness}
\pgfpathmoveto{\pgfqpoint{0pt}{0pt}}
\pgfpathlineto{\pgfpoint{\mcSize+\mcThickness}{\mcSize+\mcThickness}}
\pgfusepath{stroke}
}}
\makeatother

\newcommand\bovermat[2]{%
  \makebox[0pt][l]{$\smash{\overbrace{\phantom{%
    \begin{matrix}#2\end{matrix}}}^{\text{#1}}}$}#2}
\renewcommand{\t}[1]{ \tilde{#1} }

\newcommand{\mL}{\mathcal{L}}

\usepackage{pdfpages}
\usepackage{autobreak}
\numberwithin{equation}{section}

\author[a]{Yacov-Nir Breitstein,}
\author[b]{Adar Sharon}

\affiliation[a]{\textit{Department of Particle Physics and Astrophysics, Weizmann Institute of Science,}

\textit{Rehovot 7610001, Israel}}
\affiliation[b]{Simons Center for Geometry and Physics, SUNY, Stony Brook, NY 11794, USA}

\title{Hunting 3d $\mathcal{N}=1$ SQED in the $\epsilon$-expansion}

\emailAdd{yacov-nir.breitstein@weizmann.ac.il}			 
\emailAdd{asharon@scgp.stonybrook.edu}    

\abstract{It was recently shown that $3d$ $\mathcal{N}=1$ supersymmetric Wess-Zumino models can be studied in the $\epsilon$-expansion by analytically continuing the number of fermionic degrees of freedom to be half-integer. In this work we study the extension of this strategy to gauge theories. We consider $U(1)$ gauge theories with $N_g$ neutral Majorana fermions $\chi_a$, $N_f$ charge-1 bosons $\phi_i$ and $N_f\times N_g$ charge-1 Dirac fermions $\psi_{ia}$ in the $d=4-2\epsilon$ expansion. Analytically continuing to $N_g=\frac12$ schematically matches the Lagrangian and matter content of $3d$ $\mathcal{N}=1$ SQED, and we check whether this match can be made rigorous. We compute anomalous dimensions of $\chi_a$ up to two loops and of meson operators up to one loop at the fixed points, and compare to expectations from SUSY.  While we find obstructions to SUSY at small $N_f$, at large $N_f$ the observables approach the expected values at a SUSY fixed point. This may allow for checks of $3d$ $\mathcal{N}=1$ IR dualities between gauge theories.}

\newcommand\beq{\begin{equation}}
\newcommand\eeq{\end{equation}}

\begin{document}

\maketitle

% \tableofcontents{}

\section{Introduction}

In trying to solve strongly-coupled quantum field theories, physicists have become increasingly creative in attempts to define new parameters that can be used to perform perturbation theory. Some of the more famous examples include analytic continuation in the number of dimensions through the $\epsilon$-expansion, and analytic continuation in the number of degrees of freedom through the large-$N$ expansion.

However, some theories are not amenable to these methods. In particular, there seems to be a fundamental obstruction to studying $3d$ theories with $\mathcal{N}=1$ supersymmetry (SUSY) in the $\epsilon$-expansion: $3d$ $\mathcal{N}=1$ SUSY consists of 2 supercharges, while the minimal number of supercharges in $4d$ is 4. As a result it is not clear which $4d$ theory one should study in $d=4-2\epsilon$ in order to find $\mathcal{N}=1$ SUSY in $d=3$. Relatedly, a $3d$ Majorana fermion has 2 real components, while the smallest spinor representation in $4d$ has 4 real components, so that this issue arises even in the absence of SUSY.

A surprising solution was proposed for $3d$ $\mathcal{N}=1$ Wess-Zumino (WZ) models in \cite{Fei:2016sgs,ThomasSeminar}. These theories consist of interacting scalar multiplets, whose components include a real boson $\phi$ and a Majorana fermion $\psi$. As a first step we consider a more general (non-SUSY) class of theories with an additional fermion flavor index $a=1,...,2N_g$.
For $N_g=\frac12$ we recover the original SUSY theory, but we will instead focus on these theories for $N_g\in \mathbb{Z}$. For integer $N_g$ there is no obstruction to studying the theory in the $\epsilon$-expansion from $4d$, since we can repackage the $2N_g$ Majorana fermions into $N_g$ Dirac fermions. One can then compute all beta functions and anomalous dimensions in $d=4-2\epsilon$, which will be analytic in $N_g$. Eventually we set $N_g=\frac12$ in these equations, which should hopefully correspond to our target $3d$ $\mathcal{N}=1$ theory. This strategy was tested for the $3d$ $\mathcal{N}=1$ Ising model in \cite{Fei:2016sgs} and for additional $3d$ $\mathcal{N}=1$ theories in \cite{Liendo:2021wpo,Zerf:2017zqi,Benini:2018bhk,Benini:2018umh,Prakash:2023koy}, and the results suggest that it is indeed successful in approximating these $3d$ theories.

In this paper we attempt to extend this method to $3d$ $\mathcal{N}=1$ gauge theories. This is especially interesting due to the recent dualities that have been proposed between $3d$ $\mathcal{N}=1$ SUSY theories \cite{Jain:2013gza,Choi:2018ohn,Gaiotto:2018yjh,Dey:2019ihe,Gomis:2017ixy,Eckhard:2018raj,Benini:2018umh,Benini:2018bhk,Bashmakov:2018wts,Inbasekar:2015tsa,Aharony:2019mbc,Sharon:2020xod}, since this method can be used to check them. It also has applications to better understanding moduli spaces in CFTs from first principles, possibly allowing an extension of some of the calculations from \cite{Cuomo:2024vfk,Cuomo:2024fuy} to gauge theories. In this paper we will focus on $3d$ $\mathcal{N}=1$ SQED and leave nonabelian gauge theories for future work. 

There is an additional complication in studying SUSY gauge theories compared to WZ theories in the $\epsilon$-expansion. Supersymmetry is explicitly broken in gauge theories when using dimensional regularization, since the gaugino and gauge field have a different number of degrees of freedom in general $d$. As a result, unlike in the WZ case, we do not expect SUSY to be present for all $\epsilon$ when we set $N_g=\frac12$ in our gauge theory, and instead we must hope that SUSY appears when approaching $3d$. We are thus forced to consider a general $4d$ Lagrangian which is consistent with the expected global symmetries but is not constrained by expectations from SUSY. Then we must look for a fixed point which approaches the SUSY one as we take $\epsilon\to \frac12$. We emphasize that since SUSY is explicitly broken for general $\epsilon$, when we set $\epsilon=\frac12$ we do not expect to find exact SUSY results at each order in $\epsilon$, but only results which approximate the expectations from SUSY and improve as we add higher loop corrections.\footnote{One might be tempted to use dimensional reduction \cite{Siegel:1979wq} instead of the usual $\epsilon$-expansion. However this will not lead to $3d$ $\mathcal{N}=1$ SUSY, since we are not starting with a SUSY theory in 4d. Specifically, the gauge field will remain four-dimensional even when we set $\epsilon=\frac12$ and so it does not have the correct number of degrees of freedom for a $3d$ $\mathcal{N}=1$ vector multiplet.}

With this complication in mind, we consider general $4d$ theories with a $U(1)$ gauge field, neutral Majorana fermions $\chi_a$ (the ``gauginos''), complex charge-1 bosons $\phi_i$ and charge-1 Dirac Fermions $\psi_{ia}$ with $i=1,...,N_f$ and $a=1,...,N_g$, preserving an $SU(N_f)\times SO(N_g)$ symmetry. We will study these theories for general $N_f,N_g$ in the $\epsilon$-expansion, and find a fixed point for all integer $N_f,N_g$ where we compute the anomalous dimensions of $\chi_a$ and of meson operators $\bar{\phi_i}\phi_j$ and $\bar\phi_i\psi_{ja}$. We emphasize that our results for these theories for integer $N_f,N_g$ are new, regardless of our attempt to study $3d$ SQED.

We then set $N_g=\frac12$, where the matter content reduces to that of $3d$ $\mathcal{N}=1$ SQED, and look for a fixed point which approaches SQED as we take $\epsilon\to \frac12$. If the $3d$ fixed point has $\mathcal{N}=1$ SUSY, we expect the dimension of $\chi_a$ to approach $3/2$ as $\epsilon\to \frac12$, and the dimensions of mesons to approach the equality $\Delta_{\bar\phi\psi}=\Delta_{\bar\phi\phi}+\frac12$ (in each irreducible representation appearing in the product of the fields).

We will show that for $N_g=\frac12$, there is no unitary fixed point for $N_f<6$, but for $N_f\geq 6$, two unitary fixed points appear. As a result, we do not find hints of SUSY at small $N_f$. However, as we increase $N_f$ our results do seem to indicate the appearance of a SUSY fixed point. Specifically, 
the anomalous dimension of the ``gaugino'' $\chi_a$ is 
\begin{equation}
    \gamma_\chi=2\frac{N_f+3 }{2 N_f+7}\epsilon\;,
\end{equation}
which is very close to the expected result $\gamma_\chi=\frac12$ when we set $\epsilon=\frac12$, and approaches it more and more as we increase $N_f$. The anomalous dimensions of meson operators will also agree with the expectation from SUSY at large $N_f$, as they behave as $O\left( \frac{1}{N_f}\right)$ in the large $N_f$ expansion (these results also agree with large-$N_f$ results \cite{Benvenuti:2019}). The results at large $N_f$ are especially encouraging because it is not clear that the $\epsilon$-expansion can be trusted at small $N_f$, since by explicit calculations we find that  the 1-loop and 2-loop contributions to the gaugino anomalous dimensions are of the same order of magnitude at $\epsilon=1/2$. 

The fact that for small $\epsilon$ a unitary fixed point exists for $N_g\in\mathbb{Z}$ but not $N_g=\frac12$ suggests another way to approach the expected $3d$ fixed point: we can perform a combined $\epsilon$-expansion and analytic continuation of $N_g$ by setting $N_g=1-\epsilon$. We will call this the revised $\epsilon$-expansion.\footnote{Such a combined expansion in $\epsilon$ and the number of degrees of freedom has been used before, most notably in the method of dimensional reduction \cite{Siegel:1979wq}, although our motivation here is different.} At one loop this method coincides with setting $N_g=1$, but at two loops it leads to a unitary fixed point at small $\epsilon$ which we can study as it approaches $\epsilon=\frac12$. However, we do not approach a SUSY fixed point using this method, even at large $N_f$, since the meson anomalous dimensions do not agree with the expectations from SUSY.

We thus find encouraging results at large $N_f$, but inconclusive results at small $N_f$. There are various other ways to proceed in the hunt for $3d$ $\mathcal{N}=1$ gauge theories in the $\epsilon$-expansion. First one can extend the meson calculation to two-loop to see if at large $N_f$ the result still agrees with the expectations from SUSY. Next, a similar analysis can be done for any IR-free gauge theory, and in particular one can apply it to $4d$ QCD for large enough values of $N_f$ in order to look for $3d$ $\mathcal{N}=1$ SQCD. Finally, one can also try to approach the SUSY fixed point another way, while maintaining the expected results from SUSY for all $\epsilon$. Specifically, the number of degrees of freedom in the vector multiplet can be made to match if we also add $N_\varphi$ neutral scalars $\varphi_a$ and eventually set $N_\varphi=2\epsilon-1$. Since $N_\varphi$ is negative for $3<d<4$, this ``cancels'' the extra degrees of freedom from the gauge field, and one can hope that as a result SUSY is preserved for all $\epsilon$. However, if we preserve an $SO(N_\varphi)$ symmetry, these scalars only couple via quartic interactions with the other scalars in the theory, and so they do not affect our results for the anomalous dimension of the gaugino at one-loop, and so the results cannot agree with the SUSY expectations for all $\epsilon$. Instead one has to find a way to introduce such scalars and also couple them to the fermions via Yukawa interactions.

\section{Setup}

In this section we explicitly write down the $4d$ Lagrangian we will be studying, and motivate it from the $3d$ $\mathcal{N}=1$ perspective. Our $4d$ conventions follow \cite{Luo:2002ti}, and our choice for the $4d$ gamma matrices is
\begin{equation}
    \gamma^0 = \begin{pmatrix}
        0 & \identity\\
        \identity & 0
    \end{pmatrix}\;,
    \qquad \gamma^i = \begin{pmatrix}
        0 & \sigma^i \\
        -\sigma^i & 0
    \end{pmatrix}\;,
\end{equation}
with $\sigma^i$ the Pauli matrices, so that the $4d$ Majorana condition for a four-component spinor $\tilde\chi$ is 
\begin{equation}
    \t{\chi}^{*}=\gamma^{3}\t{\chi}\;.
\end{equation}
Our $3d$ $\mathcal{N}=1$ conventions follow \cite{Gates:1983nr}.

\subsection{$3d$ $\mathcal{N}=1$ SQED}\label{sec:SQED}

$3d$ $\mathcal{N}=1$ SQED consists of a $U(1)$ vector multiplet (consisting of a vector field $A_\mu$ and Majorana gaugino $\chi$) coupled to $N_f$ complex scalar multiplets (consisting of complex Bosons $\phi_i$ and Dirac fermions $\psi_i$), and vanishing Chern-Simons level. The Lagrangian is
\begin{equation}\label{eq:3d_SQED}
    -\frac12f^{\alpha\beta}f_{\alpha\beta}+\chi^\alpha i\partial_{\alpha}^{\;\beta}\chi_\beta+\bar{\psi}_i^\alpha\left(i \partial_\alpha{ }^\beta+eA_\alpha{ }^\beta\right) \psi_{i\beta}+\bar{\phi}_i\left(\partial_{\alpha \beta}-i eA_{\alpha \beta}\right)^2 \phi_i+\left(ie \bar{\psi}_i^\alpha \chi_{\alpha} \phi_i+h . c .\right)\;.
\end{equation}
Here $f$ is the field strength, $\alpha=1,2$ are spinor indices and $i=1,..,N_f$ is a flavor index. The parity anomaly forces $N_f\in 2\mathbb{Z}$. The continuous symmetries of this theory are $U(1)_T\times SU(N_f)$, and it also has an exact moduli space which is protected by a $\mathbb{Z}_2$ R-symmetry \cite{Gaiotto:2018yjh}. At $N_f=2$, this theory is conjectured to be dual \cite{Gremm:1999su,Gukov:2002es,Gaiotto:2018yjh,Benini:2018bhk} to a WZ theory with a complex $SU(2)$ doublet $u^a$ with $U(1)$ charge 1 and a real $SU(2)$ adjoint $R$ with superpotential $W=\bar u^a R_{ab}u^b$.

In general we expect $U(1)$ gauge theories to flow to CFTs in the IR. This can be shown rigourously at large $N_f$ \cite{PhysRevLett.64.721,PhysRevLett.60.2575}, and the common lore is that this is true also at small $N_f$. In the specific case of SQED there is a stronger argument -- adding a supersymmetric mass term for the matter fields, we find a phase transition at $m=0$ between a Higgsed phase and a free photon phase. Since the two phases are supersymmetric, the vacuum has zero energy, and so the phase transition between them has to be of second order (or higher), and so we expect a CFT at $m=0$.\footnote{This argument assumes that there are no additional phases at strong coupling which break SUSY.} This is the CFT we would like to study in this paper.

\subsection{The $4d$ Lagrangian}

We would like to study $3d$ $\mathcal{N}=1$ SQED in the $\epsilon$-expansion, but we are faced with the obstacle described in the introduction, where there is no obvious $4d$ candidate we can use. For WZ theories, a solution was proposed in \cite{Fei:2016sgs,ThomasSeminar}, which we now review. Suppose we would like to study a general $3d$ $\mathcal{N}=1$ WZ model, whose Lagrangian takes the form 
\begin{equation}
    \mL=\frac12(\partial_\mu\phi_i)^2-\frac{i}2\psi_i^\alpha\partial_{\alpha\beta}\psi_i^\beta-\frac{i}{2}\partial_i\partial_jW\psi^\alpha_i\psi_{\alpha j}-\frac12(\partial_iW)^2\;,
\end{equation}
where $W(\Psi_i,\bar\Psi_i)$ is the superpotential. We instead consider a (non-SUSY) theory with an additional flavor index $a=1,...,2N_g$ for the fermions, and Lagrangian
\begin{equation}\label{eq_eps_ep}
     \mL=\frac12(\partial_\mu\phi_i)^2-\frac{i}2\psi_{ai}^{\alpha}\partial_{\alpha\beta}\psi_{ai}^\beta-\frac{i}{2}\partial_i\partial_jW\psi^\alpha_{ai}\psi_{\alpha a j }-\frac12(\partial_iW)^2\;.
\end{equation}
Setting $N_g=\frac12$ reproduces the SUSY theory, but we will instead consider the theory for integer $N_g$, where we can repackage the $3d$ two-component Majorana fermions into $N_g$ $4d$ four-component Majorana fermions, allowing us to study the theory in $4d$ and in the $\epsilon$-expansion using standard techniques. Eventually we analytically continue to $N_g=\frac12$, which should correspond to our target $3d$ $\mathcal{N}=1$ theory.

Now we can try to extend this method to SQED. The matter content and interactions of SQED in equation \eqref{eq:3d_SQED} can be reproduced by starting in $4d$ with the same bosonic matter content but with Dirac fermions $\psi_{ia}$ and Majorana fermions $\chi_a$ (where we have introduced an additional flavor index $a=1,...,N_g$), and including the Yukawa interaction
\begin{equation}\label{eq:yukawa}
    \bar{\psi}_{ia} \chi_a \phi_i\;.
\end{equation}
This setup naively reproduces the matter content and interactions of \eqref{eq:3d_SQED} when we set $N_g=\frac12$. 

However, as mentioned above, SUSY is explicitly broken in gauge theories when using dimensional regularization, and so we have to take into account all possible interactions consistent with the global symmetries of the system. In particular, we are forced to also include a general symmetry-preserving quartic interaction $(|\phi_i|^2)^2$ as well. 

We will thus be studying the general $4d$ Lagrangian
\begin{equation}
\begin{split}
    \mathcal{L}&=-\frac{1}{4}F^{\mu\nu}F_{\mu\nu}+D_{\mu}\phi_{i}^{*}D^{\mu}\phi_{i}+\overline{\t{\psi}}_{ia}i\cancel{D}\t{\psi}_{ia}+\frac{i}{2}\overline{\t{\chi}}_{a}\cancel{\partial}\t{\chi}_{a}-\frac{1}{4}\lambda_c\phi_{i}^{*}\phi_{i}\phi_{j}^{*}\phi_{j}-\left(y\phi_{i}\overline{\t{\psi}}_{ia}\t{\chi}_{a}+h.c.\right)\;.
    \end{split}
\end{equation}
We will be using the results of \cite{Luo:2002ti} (see also \cite{Machacek:1983tz,Machacek:1983fi,Machacek:1984zw}) for the beta functions and anomalous dimensions, and so it is useful to write this explicitly in 2-component notation and in terms of real scalars. Decomposing the complex scalar into $\phi_i=\frac{1}{\sqrt{2}} \left( \phi_i^{re}+i\phi_i^{im} \right)$ and the Fermions into two-component fermions
\begin{equation}
    \tilde{\psi}_{ia} =
    \begin{pmatrix}
        \psi_{Lia} \\
        \psi_{Ria}
    \end{pmatrix}\;,\quad
    \tilde{\chi}_a =
    \begin{pmatrix}
        i\sigma^2 \chi_a^* \\
        \chi_a
    \end{pmatrix}\;,
\end{equation}
we find the Lagrangian
\begin{equation}\label{eq:Lagrangian_translated}
    \mathcal{L}	=-\frac{1}{4}F^{\mu\nu}F_{\mu\nu}+\frac{1}{2}D_{\mu}\phi_{p}D^{\mu}\phi_{p}+\psi_{ia}^{\dagger}i\sigma^{\mu}\partial_{\mu}\psi_{ia}+i\chi_{a}^{\dagger}\overline{\sigma}^{\mu}\partial_{\mu}\chi_{a}-\frac{1}{4!}\lambda_{\boldsymbol{p}}\phi_{p}\phi_{q}\phi_{r}\phi_{s}-\frac{1}{\sqrt{2}}\left[y_{i}^{p}\phi_{p}\psi_{ia}\chi_{a}+h.c.\right]\;.
\end{equation}
Let us explain the notation. $\phi_p$ runs over all real scalars, $\psi_{ia}$ runs over all charged Weyl fermions and $\chi_a$ runs over all neutral fermions. Specifically:
\begin{equation}
\begin{split}
    \{\phi_p\}_{p=1}^{2N_f}&=\left(\phi_{1}^{re},\phi_{1}^{im},\dots,\phi_{N_f}^{re},\phi_{N_f}^{im}\right)\;,\\
    \{\psi_{ia}\}_{i=1,a=1}^{2N_f,N_g}&=\left( \psi_{L11}, \psi_{R11}, \psi_{L21}, \dots, \psi_{L N_f N_g}, \psi_{R N_f N_g}\right)\;,\\
    \{\chi_a\}_{a=1}^{N_g}&=
    \left( \chi_1, \dots, \chi_{N_g} \right)\;.
\end{split}
\end{equation}

For concreteness we explicitly write down the couplings in the case $N_f=2$ in the notation of \cite{Luo:2002ti}. The quartic coupling is 
\begin{equation}
    \lambda_{pppp}=\lambda_0\;,\qquad \lambda_{pprr}=\lambda_0/3\;,\qquad \lambda_0\equiv6\lambda_c,
\end{equation}
plus permutations.
The Yukawa coupling in this notation is:

\begin{equation}
    y_i^p = \begin{cases}
                -y, & p=i \text{ or } p=i-1 \text{ and } p \text{ odd,} \\
                iy, & p=i+1 \text{ and } p \text{ even,} \\
                -iy, & p=i \text{ and } p \text{ even,} \\
                0, & \text{otherwise}.
        \end{cases}
\end{equation}

To use the results of \cite{Luo:2002ti}, we need to bring the Yukawa terms to the form used there:

\begin{equation}
    \mathcal{L}\supset -\frac12 \left( Y^p_{jk} \psi_j \psi_k \phi_p + h.c. \right).
\end{equation}
To do this, we need to treat the Yukawa coupling as a matrix in the fermionic indices, which include both the matter fermions and gauginos. We can then order it primarily by the $a=1,\dots,N_g$ indices, and find that $Y^p$ is block diagonal for every $p$, consisting of repeated identical blocks $Y^{(0)p}$. It can thus be written as:
\begin{equation}
    Y	=\begin{pmatrix}Y^{(0)}\\
 & Y^{(0)}\\
 &  & \cdots\\
 &  &  & Y^{(0)}
\end{pmatrix}_{a=1}^{N_{g}}\;.
\end{equation}
The indices inside each block are the flavor indices $j=1,\dots,(2N_f+1)$, and the order is $\left(\chi_a,\psi_{L1a}, \psi_{R1a},\dots, \psi_{RN_fa} \right)$. For $N_f=2$ the blocks are:
\begin{equation}
    \begin{split}
        Y^{(0)1}	=-\frac{y}{\sqrt{2}}\begin{pmatrix} & 1 & 1 & 0 & 0\\ 1\\ 1\\ 0\\ 0
\end{pmatrix}\;,\qquad  & 
Y^{(0)2}	=\frac{y}{\sqrt{2}}\begin{pmatrix} & i & -i & 0 & 0\\i\\-i\\0\\0
\end{pmatrix} \;,\\
Y^{(0)3}	=-\frac{y}{\sqrt{2}}\begin{pmatrix} &  &  & 1 & 1\\
\\ \\1\\1
\end{pmatrix} \;,\qquad  & Y^{(0)4}	=\frac{y}{\sqrt{2}}\begin{pmatrix} &  &  & i & -i\\ \\ \\i\\-i
\end{pmatrix}\;.
    \end{split}
\end{equation}
This can be generalized for general $N_f$ in a straightforward manner. 
To relate these new couplings to the previous ones, we consider any $a$-block of $Y^p_{jk}$, $Y^{(0)p}_{jk}$. Within this block, and for $j,k>1$, we have:

\begin{equation}
        Y^{(0)p}_{j1} = Y^{(0)p}_{1j} = \frac1{\sqrt{2}} y_{j-1}^p\;,\qquad        Y^{(0)p}_{11} = Y^{(0)p}_{jk} = 0\;.
\end{equation}

\section{Results}

We now present our results. We will be particularly interested in the case $N_g=\frac12$ and check whether we approach a SUSY fixed point in $3d$, although the results for integer $N_g$ are also new. We will compute several observables to try to gauge whether we approach a SUSY fixed point. The first is the two-loop anomalous dimension of the gaugino $\chi_a$, which SUSY fixes to be exactly $\gamma_\chi=\epsilon$ in $d=4-2\epsilon$ dimensions (so that its dimension is $\Delta_\chi=\frac32$ for all $d$). The second is the one-loop anomalous dimensions of scalar mesons $\phi_i^*\phi_j$ and mixed mesons $\phi_i^*\psi_j$, with $i,j$ flavor indices. SUSY fixes $\Delta_{\phi^*\psi}=\Delta_{\phi^*\phi}+\frac12$ for both the adjoint and singlet contraction of the flavor indices.

\subsection{Beta functions and fixed points}

We present the general results for the beta functions at two loops. We denote a beta function for a coupling $g$ in a loop expansion as
\begin{equation}
    \beta^g=\beta_0^g+\frac{1}{(4\pi)^2}\beta_1^g+\frac{1}{(4\pi)^4}\beta_2^g+...\;,
\end{equation}
and similarly for an anomalous dimension.

The gauge coupling beta functions are
\begin{equation}
\begin{split}
    \beta_0^e&=-\epsilon e\;,\\
    \beta_1^e&=\frac{1}{3}\left(4 N_g+ 1\right)e^3N_f\;,\\
    \beta_2^e&=\left(8 e^2 N_g+8 e^2-4 N_g y^2\right)e^3N_f\;.
\end{split}
\end{equation}
The Yukawa coupling beta functions are
\begin{equation}
\begin{split}
    \beta_0^y=&-\epsilon y\;,\\
    \beta_1^y=&\left[\left(N_f+2 N_g +\frac{5}{2}\right)y^2-3 e^2\right]y\;,\\
    \beta_2^y=&\left[\left(\frac{4}{3} N_fN_g+\frac{4 N_f}{3}-6\right)e^4+\left( \frac{17}{2}N_f+5 N_g+\frac{19}{4}\right)e^2y^2+\right.\\
    &\left.\left(-6 N_fN_g -\frac{5 N_f}{2}-9 N_g-1\right)y^4
   +\frac{\lambda_0 ^2}{6}-4 \lambda_0  y^2\right] y\;.
\end{split}
\end{equation}
The quartic beta functions are
\begin{equation}
\begin{split}
    % \beta_0^{\lambda_0}=&-2\epsilon\lambda_0\;,\\
    % \beta_1^{\lambda_0}=&36 e^4-12 e^2 \lambda_0 +\frac{2}{3}\left(4+N_f\right)\lambda_0^2-48 N_g y^4+8 \lambda_0  N_g y^2\;,\\
    % \beta_2^{\lambda_0}=&-\frac{14 \lambda_0 ^3}{3}-6   N_f^2 N_g \lambda_0 y^4-2N_f \lambda_0^3 +96 N_f N_g y^6-7   N_f N_g\lambda_0 y^4-\frac{8}{3} N_f N_g\lambda_0 ^2  y^2+336 N_g y^6\\
    % &+48  N_g\lambda_0  y^4-\frac{32}{3} N_g \lambda_0^2 y^2
    % +e^2 \left(\frac{40 \lambda_0 ^2}{3}+\frac{16 \lambda_0 ^2 N_f}{3}+20 N_g\lambda_0  y^2\right)\\
    % &+e^4 \left(58 \lambda_0 +\frac{142 \lambda_0  N_f}{3}+\frac{40 \lambda_0  N_f N_g}{3}-48 N_g y^2\right)
    % +e^6 (-128 N_f N_g-56 N_f-360)\;. \\
% 
    \beta_0^{\lambda_0}=&-2\epsilon\lambda_0\;,\\
    \beta_1^{\lambda_0}=&36 e^4-12 e^2 \lambda_0 +\frac{2}{3}\left(4+N_f\right)\lambda_0^2-48 N_g y^4+8 \lambda_0  N_g y^2\;,\\
    \beta_2^{\lambda_0}=&-\left( \frac{14}{3}+2N_f \right)\lambda_0 ^3 +\left(48 - 7 N_f - 6 N_f^2  \right) N_g \lambda_0 y^4 - \frac{8}{3} \left(  N_f +4\right) N_g \lambda_0 ^2  y^2 \\
    & +\left( 96 N_f +336 \right) N_g y^6 + \left(\frac{40 \lambda_0 ^2}{3}+\frac{16 \lambda_0 ^2 N_f}{3}+20 N_g\lambda_0  y^2\right) e^2\\
    &+ \left(58 \lambda_0 +\frac{142 \lambda_0  N_f}{3}+\frac{40 \lambda_0  N_f N_g}{3}-48 N_g y^2\right)e^4
    - (128 N_f N_g+56 N_f+360) e^6 \;.    
\end{split}
\end{equation}
Finally, the gaugino anomalous dimension is
\begin{equation}
    \begin{split}
    \gamma_1^\chi=&N_fy^2\;,\\
    \gamma_2^\chi=&\frac{17}{2}  N_f e^2y^2-3 N_f N_g y^4-\frac{N_f y^4}{4}\;.
\end{split}
\end{equation}

We can now look for fixed points. We solved for the value of all couplings at the fixed points up to two loops, but these are too long to present here. Instead we present only the 1-loop results for the fixed-point couplings.
At one-loop we find the fixed points
\begin{equation}\label{eq:fixed_points}
    \begin{split}
        e_*^2=&\frac{48 \pi ^2}{4 N_f N_g+N_f}\epsilon \;,\\
        y_*^2=&\frac{32 \pi ^2 (4 N_f N_g+N_f+9)}{N_f (4 N_g+1) (2 N_f+4 N_g+5)}\epsilon \;,\\
        \lambda_{0*}^{\pm}=&\frac{24\pi^2}{N_f (N_f+4) (4 N_g+1) (2 N_f+4 N_g+5)}(\lambda_{01}\pm \sqrt{\lambda_{02}})\epsilon \;,
    \end{split}
\end{equation}
where
\begin{equation}
    \begin{split}
        \lambda_{01}=&N_f (N_f (8 N_g+2)-16 (N_g-1) N_g+41)+90\;,\\
        \lambda_{02}=&256 N_f^2 N_g^4+256 N_f^2 (7 N_f+30) N_g^3+16 \left((N_f (2 N_f+53)+132) N_f^2+432\right) N_g\\
        &+32 (N_f (N_f (2 N_f (N_f+19)+383)+954)-432) N_g^2+(2 N_f+5)^2 ((N_f-180) N_f-540)\;.
    \end{split}
\end{equation}
The existence of a unitary fixed point requires $\lambda_{02}>0$. For $N_g\in\mathbb{N}$, the fixed points are unitary for all $N_f$. However, for $N_g=\frac12$, the two fixed points merge at $N_f=N_f^{c}\approx 5.5$ and go into the complex plane for $N_f<N_f^{c}$, and so there exists a unitary fixed point only for $N_f>N_f^{c}$. These facts remain true also at two loops. 

We can also repeat this analysis for the revised $\epsilon$-expansion, where $N_g=1-\epsilon$. In this case the fixed points are unitary for all $N_f$ for $N_g\in\mathbb{Z}$ and also for $N_g=\frac12$. However, only the coupling $\lambda_{0*}^+$ is positive, meaning the theory is only perturbatively stable at the respective fixed point. At one-loop, the fixed-point values for the couplings match those of the standard $\epsilon$-expansion \eqref{eq:fixed_points} when we set $N_g=1$, since the revised $\epsilon$-expansion only affects results at higher orders.

\subsection{Gaugino anomalous dimension}

The $\chi_a$ are gauge-invariant and so we can meaningfully compute their anomalous dimensions. Since we expect $\chi_a$ to become the gaugino for $3d$ SQED, we also call it the gaugino. If the fixed point we found in the $\epsilon$-expansion indeed becomes SQED in $3d$, we expect $\gamma_\chi$ to approach $\frac12$ as $\epsilon\to \frac12$.

At one loop, the gaugino anomalous dimension is
\begin{equation}
    \gamma_\chi=2\frac{  4 N_f N_g+N_f+9}{(4 N_g+1) (2 N_f+4 N_g+5)}\epsilon\;.
\end{equation}
Plugging in $N_g=\frac12$, we find
\begin{equation}
    \gamma_\chi=2\frac{N_f+3 }{2 N_f+7}\epsilon\;.
\end{equation}
At large $N_f$, $\gamma_\chi$ approaches the expected value for a SUSY theory of $\gamma_\chi=\frac12$ regardless of $N_g$. The result is close to the expected value even for the smallest value of $N_f=2$, where $\gamma_\chi=5/11$ (the result is real even though the fixed point is non-unitary, since the quartic coupling does not contribute to $\gamma_\chi$ at one loop).

\begin{figure}
    \centering
    \begin{subfigure}[b]{0.45\textwidth}
         \centering
         \includegraphics[width=\textwidth]{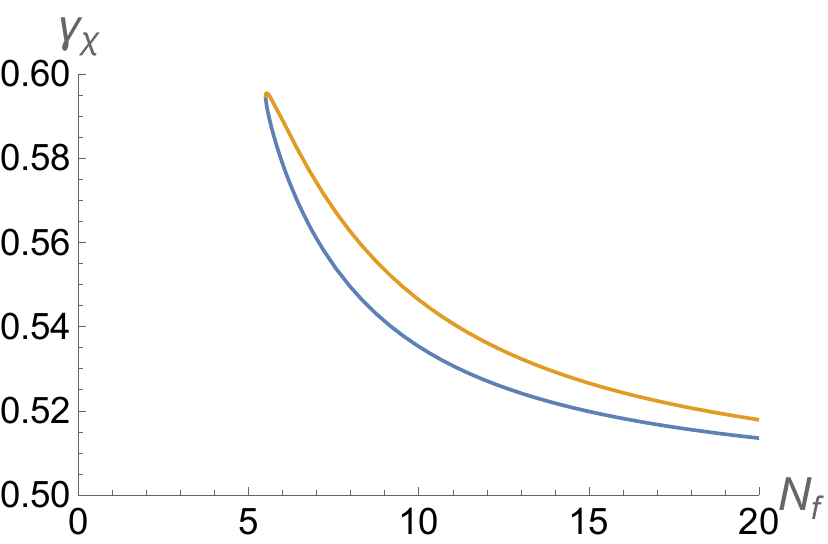}
         \caption{ }
         \label{fig:gamma_lambda_1}
     \end{subfigure}
     \hfill
     \begin{subfigure}[b]{0.45\textwidth}
         \centering
         \includegraphics[width=\textwidth]{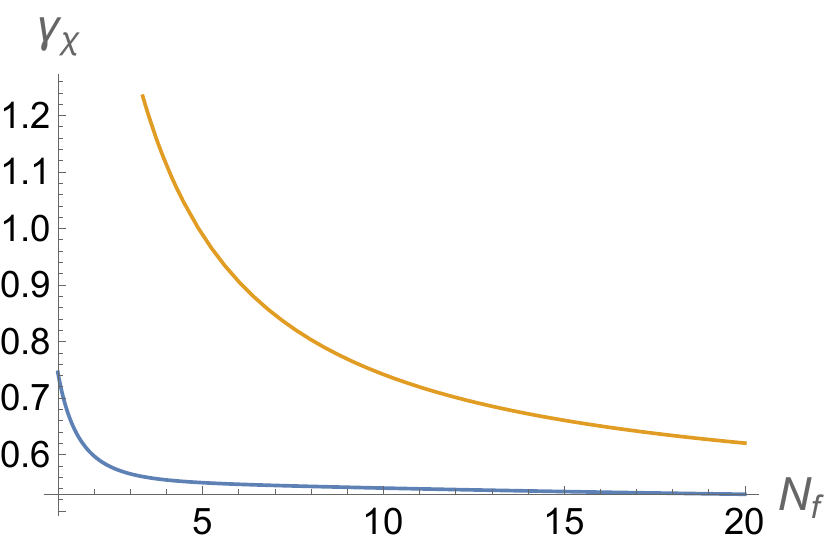}
         \caption{ }
         \label{fig:gamma_lambda_2}
     \end{subfigure}
    \caption{Two-loop gaugino anomalous dimensions, with each line representing the value at one fixed point. (a) In the standard $\epsilon$-expansion the fixed points are non-unitary for small $N_f$, but merge and become real at $N_f\approx 5.5$, and at large $N_f$ approach the expected SUSY value $\gamma_\chi=\frac12$. (b) In the revised $\epsilon$-expansion, the fixed points are unitary for all $N_f$ and again approach the SUSY value as $N_f$ increases.}
    \label{fig:gamma_lambda}
\end{figure}

We spare the reader the expression for $\gamma_\chi$ at two loops. The same physics applies also at this order: it is still true that there exists a unitary fixed point for all $N_g,N_f\in\mathbb{Z}$, and that at $N_g=\frac12$ the fixed point is only unitary for $N_f>N_f^c\approx 5.5$. At large $N_f$, $\gamma_\chi$  still approaches the SUSY value $\gamma_\chi=\epsilon+O(1/N_f)$ regardless of $N_g$. We present the results for $N_g=\frac12$ in figure \ref{fig:gamma_lambda_1}.

We can also perform the revised $\epsilon$-expansion, where we set $N_g=1-\epsilon$. In this case the fixed points are unitary for all $N_f$, and again $\gamma_\chi$ approaches the SUSY value at large $N_f$. At one loop we have:

\begin{equation}
    \gamma_\chi=\frac25 \frac{5N_f+9}{2N_f+9}\epsilon\;.
\end{equation}
At two loops the results are presented in figure \ref{fig:gamma_lambda_2}.

The results seem to suggest that we do not approach a SUSY fixed point for small $N_f$, but that at large $N_f$ we may be approaching one. It is not clear that a two-loop analysis is enough in order to rule out SUSY at small $N_f$, since the 1-loop and 2-loop contributions to the anomalous dimensions are of the same order of magnitude, and so the two-loop expansion does not seem like a good approximation. It is possible that our approximation is only valid at large-$N_f$ where additional contributions are suppressed, which would indicate that a SUSY fixed point is indeed reached.

\subsection{Mesons}

In this section we analyze the anomalous dimensions of meson operators. We will focus on two types of mesons operators:
\begin{equation}
    \bar\phi_i\phi_j\;,\qquad \bar\phi_i\psi_{ja}\;,
\end{equation}
with $i=1,...,N_f$ a flavor index and $a=1,..,N_g$ the auxiliary index. We are interested in their one-loop anomalous dimensions.

To compute \textit{e.g.}~the anomalous dimension of the scalar meson $\phi_i\bar\phi_j$, we compute the three-point correlator
\begin{equation}
    \langle \bar\phi_i\phi_j(p)\bar\phi_k(-p_1)\phi_l(p_2) \rangle\;.
\end{equation}
While the correlator itself is gauge-dependent, the anomalous dimension of the meson is not. Thus, we can compute in a specific gauge, and then apply the result for the latter independently of the gauge.
In the $\epsilon$-expansion, the one-loop result will take the form
\begin{equation}
    \langle \bar\phi_i\phi_j(p)\bar\phi_k(-p_1)\phi_l(p_2) \rangle=-\frac{Z_{ijkl}}{p_1^2p_2^2}\;,
\end{equation}
where
\begin{equation}
    Z_{ijkl}=(\delta_{il}\delta_{kj}+\frac{1}{\epsilon}z_{ijkl}+...)\;.
\end{equation}
Here the first term in the brackets is the tree-level contribution, and the second term is the diverging part of the one-loop result (and we ignore non-diverging terms). The anomalous dimension is then given by 
\begin{equation}
    \gamma_{\bar\phi_i\phi_j\bar\phi_k\phi_l}=\frac{\partial}{\partial \log \mu}\left(\frac{Z_{ijkl}}{Z_{il}Z_{kj}}\right)\;,
\end{equation}
where $Z_{ij}$ is the scalar wavefunction renormalization, which we recover from the anomalous dimension found in \cite{Luo:2002ti}. 
In principle we would now have to diagonalize the anomalous dimension matrix, but in practice the diagonalization is immediate since we know there should be an adjoint and a singlet meson, which we denote by $(\bar\phi\phi)_{\text{adj}}$ and $(\bar\phi\phi)_{\text{sing}}$. For the scalar and mixed mesons, the dimension is then given by
\begin{align}
    \Delta_{(\bar\phi\phi)_{\text{adj}}}=2-2\epsilon+\gamma_{(\bar\phi\phi)_{\text{adj}}}\;, \\
    \Delta_{(\bar\phi\psi)_{\text{adj}}}=\frac{5}{2} -2\epsilon+\gamma_{(\bar\phi\psi)_{\text{adj}}}\;,
\end{align}
and similarly for $(\bar\phi\phi)_{\text{sing}}$. 
If in $3d$ we recover SQED, we expect to find
\begin{equation}
    \gamma_{(\bar\phi\phi)_{\text{adj}}}=\gamma_{(\bar\phi\psi)_{\text{adj}}}\;,\qquad 
    \gamma_{(\bar\phi\phi)_{\text{sing}}}=\gamma_{(\bar\phi\psi)_{\text{sing}}}\;.
\end{equation}

We now present the results, with the details appearing in appendix \ref{app_mesons}. 
For general $N_f$ and $N_g$, the 1-loop anomalous dimensions are:
\begin{align}
    \gamma_{(\phi^*\phi)_{\text{adj}}} = & \frac{1}{8\pi^2} \left( \frac{\lambda_c}{2} -2N_gy^2 +3e^2 \right), \\
    \gamma_{(\phi^*\phi)_{\text{sing}}} = & \frac{1}{8\pi^2} \left( \frac{N_f+1}{2} \lambda_c -2N_gy^2 +3e^2 \right), \\
    \gamma_{(\phi^*\psi)_{\text{adj}}}=\gamma_{(\phi^*\psi)_{\text{sing}}} = & \frac{1}{(4\pi)^2} \left( 3e^2-\left(2N_{g}+\frac{1}{2}\right)y^2 \right).
\end{align}

\begin{figure}
    \centering
    \begin{subfigure}[b]{0.49\textwidth}
         \centering
         \includegraphics[width=\textwidth]{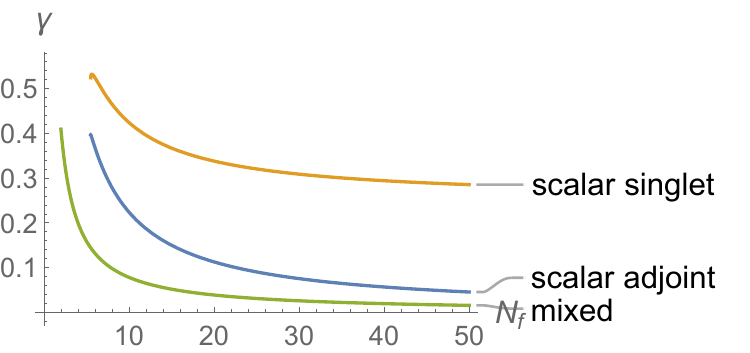}
         \caption{ }
         \label{fig:gamma_mesons+}
     \end{subfigure}
     \hfill
     \begin{subfigure}[b]{0.49\textwidth}
         \centering
         \includegraphics[width=\textwidth]{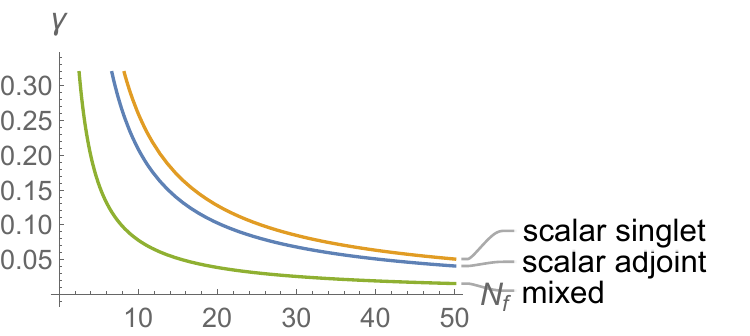}
         \caption{ }
         \label{fig:gamma_mesons-}
     \end{subfigure}
    \caption{Anomalous dimensions of the scalar and mixed mesons as a function of $N_f$, as obtained in the standard $\epsilon$-expansion. (a) In the $\lambda_{c*}^+$ fixed point, as $N_f$ grows, the anomalous dimensions of the adjoint scalar mesons and all mixed mesons tend to 0, but that of the singlet scalar meson tends to $\frac{1}{4}$. (b) In the $\lambda_{c*}^-$ fixed point, all anomalous dimensions tend to 0, in agreement with SUSY.}
    \label{fig:gamma_mesons}
\end{figure}

In the standard $\epsilon$-expansion, the results for $N_g=\frac12$ appear in figure \ref{fig:gamma_mesons}. Again, the fixed points are non-unitary for small $N_f$ and so the anomalous dimensions are complex, but become real for large enough $N_f$.
At large $N_f$, at the fixed point $\lambda=\lambda^-_{0*}$, all anomalous dimensions behave as $\sim 1/N_f$, and so at large $N_f$ our results are consistent with a SUSY fixed point, since the anomalous dimensions of scalar and mixed mesons in both representations are equal. This is also in agreement with the results of the large $N_f$ expansion done in \cite{Benvenuti:2019}. On the other hand, at the fixed point with $\lambda=\lambda_{0*}^+$, 
at large $N_f$ the anomalous dimension of the singlet contraction of $\bar\phi\phi$ tends to a finite number, while the rest are $O\left( \frac1{N_f}\right)$.
We thus conclude that only the fixed point with $\lambda^-_{0*}$ can lead to SUSY in $3d$ at large $N_f$.

\begin{figure}
    \centering
    \includegraphics[width=0.6\textwidth]{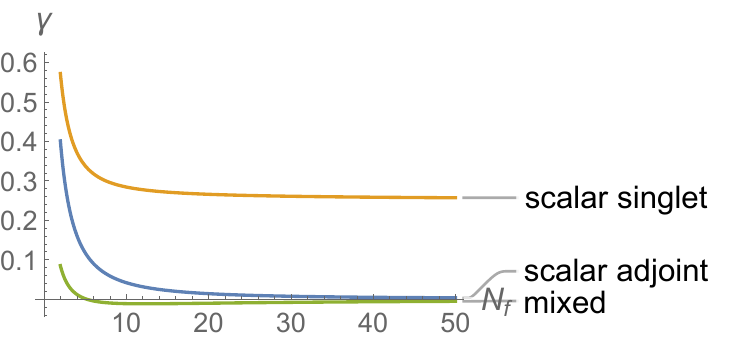}
    \hfill   
    \caption{Anomalous dimensions of the scalar and mixed mesons as a function of $N_f$, as obtained in the revised $\epsilon$-expansion. Only the fixed point with $\lambda_{c*}^+$ gives a perturbatively stable theory, and this is the one presented. As $N_f$ grows, the anomalous dimensions of the adjoint scalar mesons and all mixed mesons tend to 0, but that of the singlet scalar meson tends to $\frac{1}{4}$, so that we do not approach a SUSY fixed point.}
    \label{fig:gamma_mesons_revised}
\end{figure}

The results for the revised $\epsilon$-expansion appear in figure \ref{fig:gamma_mesons_revised}. Only the fixed point with $\lambda^+_{0*}$ leads to a perturbatively stable theory. At small $N_f$ we do have a unitary fixed point, but it does not look like a SUSY fixed point. In particular for $N_f=2$ we find
\begin{align}
    \gamma_{\left( \phi^*\phi\right)_{\text{adj}}} \approx 0.401\;, \indent
    & \gamma_{\left( \phi^*\phi\right)_{\text{sing}}} \approx 0.572\;,
    & \gamma_{\left( \phi^*\psi_\alpha\right)_{\text{all}}} \approx 0.0846\;. 
\end{align}
At large $N_f$ SUSY is violated since $\gamma_{\left( \phi^*\phi\right)_{\text{sing}}}$ approaches $1/4$ while $\gamma_{\left( \phi^*\psi_\alpha\right)_{\text{all}}} $ approaches $0$. We conclude that the revised $\epsilon$-expansion fails to lead to SUSY in $3d$.


In order to perform yet another check to see whether we approach SQED fixed point for the specific value of $N_f=2$, one can also use the proposed duality to the WZ model discussed in section \ref{sec:SQED}. The duality maps mesons to specific operators in the WZ model and by comparing our results to those obtained in the $\epsilon$-expansion for the WZ model we can see whether we approach the expected result. In agreement with the results above which indicate that we do not approach a SUSY fixed point for small $N_f$, our results are far from the expected anomalous dimensions that the duality proposes in both the standard $\epsilon$-expansion (where the fixed point isn't unitary) and the revised $\epsilon$-expansion.

\section*{Acknowledgements}

The authors would like to thank O. Aharony for many helpful discussions and for comments on a draft, and thank I. Klebanov for comments on a draft. This work was supported in part by Israel Science Foundation grant no. 2159/22, by Simons Foundation
grant 994296 (Simons Collaboration on Confinement and QCD Strings), by the Minerva
foundation with funding from the Federal German Ministry for Education and Research,
by the German Research Foundation through a German-Israeli Project Cooperation (DIP) grant "Holography and the Swampland", and by a research grant from Martin Eisenstein.

\newpage

\begin{appendix}

\addtocontents{toc}{\protect\setcounter{tocdepth}{2}}

\section{Meson calculations}\label{app_mesons}

We compute the anomalous dimensions of mesonic operators using a direct computation of correlation functions. The simplest correlation function to consider for the scalar mesons is
$\left< \left( \phi^*_i \phi_j\right)(p) \phi^*_k(-p_1) \phi_l(p_2) \right>$,
which is understood to be in terms of bare operators. The relation to renormalized operators is:
\begin{equation}
    \left< \left( \phi^*_i \phi_j\right)_B(p) \phi^*_{Bk}(-p_1) \phi_{Bl}(p_2) \right> = 
    Z_{\phi^* \phi} Z_{\phi}^2 \left< \left( \phi^*_i \phi_j\right)_R(p) \phi^*_{Rk}(-p_1) \phi_{Rl}(p_2) \right>.
\end{equation}
The convention for renormalization of fields is $\phi_B=Z_{\phi} \phi_R$ and similarly for other operators. This is different compared to the convention of e.g. \cite{Luo:2002ti} (where the fields are renormalized by $\phi_B=\sqrt{Z_{\phi}} \phi_R$).

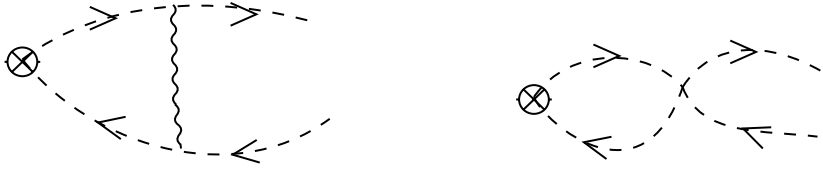
\begin{figure}
    \centering
    \tikzset{every picture/.style={line width=0.75pt}} %set default line width to 0.75pt        

\begin{tikzpicture}[x=0.75pt,y=0.75pt,yscale=-1,xscale=1]
%uncomment if require: \path (0,133); %set diagram left start at 0, and has height of 133

%Curve Lines [id:da989505248418967] 
\draw  [dash pattern={on 4.5pt off 4.5pt}]  (40,61.5) .. controls (80,31.5) and (136,28) .. (188,43) ;
%Curve Lines [id:da2841285561914615] 
\draw  [dash pattern={on 4.5pt off 4.5pt}]  (40,61.5) .. controls (71,101.5) and (143,129.5) .. (195,91.5) ;
%Straight Lines [id:da8450552435968222] 
\draw    (116,34) .. controls (117.71,35.63) and (117.75,37.29) .. (116.13,39) .. controls (114.5,40.71) and (114.54,42.37) .. (116.25,44) .. controls (117.96,45.63) and (118,47.29) .. (116.38,49) .. controls (114.76,50.71) and (114.8,52.37) .. (116.51,53.99) .. controls (118.22,55.62) and (118.26,57.28) .. (116.64,58.99) .. controls (115.01,60.7) and (115.05,62.36) .. (116.76,63.99) .. controls (118.47,65.62) and (118.51,67.28) .. (116.89,68.99) .. controls (115.27,70.7) and (115.31,72.36) .. (117.02,73.99) .. controls (118.73,75.62) and (118.77,77.28) .. (117.15,78.99) .. controls (115.52,80.7) and (115.56,82.36) .. (117.27,83.98) -- (117.29,84.57) -- (117.29,84.57) .. controls (119.15,86.02) and (119.35,87.67) .. (117.9,89.53) .. controls (116.45,91.39) and (116.66,93.04) .. (118.52,94.49) .. controls (120.38,95.94) and (120.58,97.6) .. (119.13,99.46) .. controls (117.68,101.32) and (117.88,102.97) .. (119.74,104.42) -- (120,106.5) -- (120,106.5) ;
%Curve Lines [id:da7122294810432075] 
\draw  [dash pattern={on 4.5pt off 4.5pt}]  (298,80.48) .. controls (316,54.5) and (356,55.5) .. (373,75.5) ;
%Curve Lines [id:da9138217680239233] 
\draw  [dash pattern={on 4.5pt off 4.5pt}]  (298,80.48) .. controls (303,93.5) and (353,136.32) .. (373,75.5) ;
%Curve Lines [id:da1406138388427658] 
\draw  [dash pattern={on 4.5pt off 4.5pt}]  (373,75.5) .. controls (388,51.5) and (417,52.5) .. (443,67.5) ;
%Curve Lines [id:da43252773082496] 
\draw  [dash pattern={on 4.5pt off 4.5pt}]  (373,75.5) .. controls (382,100.48) and (414,97.2) .. (441,100.5) ;
%Shape: Light Bulb [id:dp6307239031798082] 
\draw   (32.5,62.75) .. controls (32.5,58.75) and (35.86,55.5) .. (40,55.5) .. controls (44.14,55.5) and (47.5,58.75) .. (47.5,62.75) .. controls (47.5,66.75) and (44.14,70) .. (40,70) .. controls (35.86,70) and (32.5,66.75) .. (32.5,62.75) -- cycle (34.66,57.59) -- (45.34,67.91) (45.34,57.59) -- (34.66,67.91) (31,62.75) -- (32.5,62.75) (47.5,62.75) -- (49,62.75) ;
\draw   (74,35) -- (87,40.75) -- (74,46.5) ;
\draw   (159.71,113.58) -- (146.06,109.62) -- (158.18,102.18) ;
\draw   (89.65,101.74) -- (78.14,93.39) -- (92.06,90.49) ;
\draw   (145,33) -- (158,38.75) -- (145,44.5) ;
\draw   (397,52) -- (410,57.75) -- (397,63.5) ;
\draw   (328,54) -- (341,59.75) -- (328,65.5) ;
\draw   (413.44,106.48) -- (403.45,96.37) -- (417.66,95.78) ;
\draw   (334.58,111.79) -- (323.16,103.32) -- (337.11,100.57) ;
%Shape: Light Bulb [id:dp48119309953445155] 
\draw   (290.5,81.75) .. controls (290.5,77.75) and (293.86,74.5) .. (298,74.5) .. controls (302.14,74.5) and (305.5,77.75) .. (305.5,81.75) .. controls (305.5,85.75) and (302.14,89) .. (298,89) .. controls (293.86,89) and (290.5,85.75) .. (290.5,81.75) -- cycle (292.66,76.59) -- (303.34,86.91) (303.34,76.59) -- (292.66,86.91) (289,81.75) -- (290.5,81.75) (305.5,81.75) -- (307,81.75) ;

\end{tikzpicture}
    \caption{Diagrams contributing to the scalar meson renormalization function at 1-loop order. Other diagrams, involving propagator corrections, are accounted for separately through the field renormalization functions.}
    \label{fig:scalar_meson_diagrams}
\end{figure}

The $Z$ factors are chosen according to the MS scheme.
We can deduct the scalar renormalization, which for $\epsilon=\frac{1}{2}$ equals (in $R_\xi$ gauge):
\begin{equation}
    Z_\phi = 1+\frac{1}{(4\pi)^2} \left( -2N_g y^2 + (3-\xi)e^2 \right)\;.
\end{equation}
Then, the anomalous dimensions of the meson operator can be derived via the relation
\begin{equation}
    \gamma_{\phi^*\phi} = \frac{\partial \log Z_{\phi^*\phi}}{ \partial\log \mu},
\end{equation}
with $\mu$ the renormalization scale. The diagrams we need to compute are shown in figure \ref{fig:scalar_meson_diagrams}. There are additional diagrams, contributing to the external leg corrections. However, we can infer their sum from $Z_{\phi}$, which we derive from its relation to $\gamma_\phi$, which in turn we can compute from the formulas in \cite{Luo:2002ti}.

We can thus extract the renormalization function:
\begin{equation}
    Z_{(\phi^*\phi), ijkl} = \left[1-\frac{1}{\left(4\pi\right)^{2}\epsilon}\left(\frac{\lambda_c}{12}-2N_{g}y^{2}+3e^{2}\right)\right]\delta_{il}\delta_{kj}-\frac{\lambda_c}{12\left(4\pi\right)^{2}\epsilon}\delta_{ij}\delta_{kl}.
\end{equation}
This expression should be viewed as an $N_f^2\times N_f^2$ matrix, with the indices $i,j$ indicating the row and $k,l$ indicating the column. When we order the index multiple $(i,j)$ first by $i$ and then by $j$, and the columns first by $l$ and then by $k$, the renormalization matrix is of the form:
\begin{equation}
    AI_{N_f^2\times N_f^2}+\begin{pmatrix}
        B &\bovermat{$N_f$}{ 0 & \cdots & 0}  & B & 0 &\cdots& 0  & B \\
        0 & 0 &  &  & 0 &0 & &0&0 \\
        \vdots &  & \ddots  & & \vdots & &\ddots&&\vdots \\
        0 & 0 &  &  0& 0 &0 & &0&0\\
        B & 0 & \cdots &  0& B  & 0 & \cdots & 0 &B\\
        0 & 0&  & 0& 0 & 0 & &0&0\\
         \vdots & & \ddots & & \vdots &  &\ddots &&\vdots \\
        B &0 & \cdots & 0 &B &0 & \cdots &0&B
    \end{pmatrix},
\end{equation}
with $I$ the identity matrix, $A = 1-\frac{1}{\left(4\pi\right)^{2}\epsilon}\left(\frac{\lambda_c}{12}-2N_{g}y^{2}+3e^{2}\right)$ and $B = -\frac{\lambda_c}{12\left(4\pi\right)^{2}\epsilon}$.
The eigenvalues of this matrix are $A$ with a multiplicity of $N_f^2-1$ and $A+N_fB$ with a multiplicity of $1$. These correspond to the adjoint and singlet representations appearing in the product $\mathbf{N_f}\times\mathbf{\overline{N_f}}$, respectively. The anomalous dimensions are then:
\begin{align}
    \gamma_{\left( \phi^*\phi\right)_{\text{adj}}} = \frac{1}{8\pi^{2}}\left(\frac{\lambda_c}{2}-2N_{g}y^{2}+3e^{2}\right),\\
    \gamma_{\left( \phi^*\phi\right)_{\text{sing}}} = \frac{1}{8\pi^{2}}\left(\frac{N_f+1}{2}\lambda_c-2N_{g}y^{2}+3e^{2}\right).
\end{align}

\begin{figure}
    \centering
    \tikzset{every picture/.style={line width=0.75pt}} %set default line width to 0.75pt        

\begin{tikzpicture}[x=0.75pt,y=0.75pt,yscale=-1,xscale=1]
%uncomment if require: \path (0,133); %set diagram left start at 0, and has height of 133

%Curve Lines [id:da989505248418967] 
\draw    (40,61.5) .. controls (80,31.5) and (136,28) .. (188,43) ;
%Curve Lines [id:da2841285561914615] 
\draw    [dash pattern={on 4.5pt off 4.5pt}] (40,61.5) .. controls (71,101.5) and (143,129.5) .. (195,91.5) ;
%Straight Lines [id:da8450552435968222] 
\draw    (116,34) .. controls (117.71,35.63) and (117.75,37.29) .. (116.13,39) .. controls (114.5,40.71) and (114.54,42.37) .. (116.25,44) .. controls (117.96,45.63) and (118,47.29) .. (116.38,49) .. controls (114.76,50.71) and (114.8,52.37) .. (116.51,53.99) .. controls (118.22,55.62) and (118.26,57.28) .. (116.64,58.99) .. controls (115.01,60.7) and (115.05,62.36) .. (116.76,63.99) .. controls (118.47,65.62) and (118.51,67.28) .. (116.89,68.99) .. controls (115.27,70.7) and (115.31,72.36) .. (117.02,73.99) .. controls (118.73,75.62) and (118.77,77.28) .. (117.15,78.99) .. controls (115.52,80.7) and (115.56,82.36) .. (117.27,83.98) -- (117.29,84.57) -- (117.29,84.57) .. controls (119.15,86.02) and (119.35,87.67) .. (117.9,89.53) .. controls (116.45,91.39) and (116.66,93.04) .. (118.52,94.49) .. controls (120.38,95.94) and (120.58,97.6) .. (119.13,99.46) .. controls (117.68,101.32) and (117.88,102.97) .. (119.74,104.42) -- (120,106.5) -- (120,106.5) ;
%Shape: Light Bulb [id:dp6307239031798082] 
\draw   (32.5,62.75) .. controls (32.5,58.75) and (35.86,55.5) .. (40,55.5) .. controls (44.14,55.5) and (47.5,58.75) .. (47.5,62.75) .. controls (47.5,66.75) and (44.14,70) .. (40,70) .. controls (35.86,70) and (32.5,66.75) .. (32.5,62.75) -- cycle (34.66,57.59) -- (45.34,67.91) (45.34,57.59) -- (34.66,67.91) (31,62.75) -- (32.5,62.75) (47.5,62.75) -- (49,62.75) ;
\draw   (74,35) -- (87,40.75) -- (74,46.5) ;
\draw   (159.71,113.58) -- (146.06,109.62) -- (158.18,102.18) ;
\draw   (89.65,101.74) -- (78.14,93.39) -- (92.06,90.49) ;
\draw   (145,33) -- (158,38.75) -- (145,44.5) ;

\end{tikzpicture}
    \caption{Diagram contributing to the mixed meson renormalization function at 1-loop order. Other diagrams, involving propagator corrections, are accounted for separately through the field renormalization functions.}
    \label{fig:mixed_meson_diagram}
\end{figure}
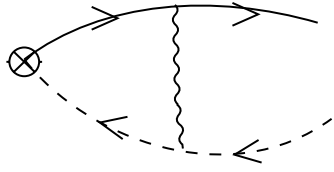

We compute the anomalous dimensions of the mixed meson operators in a similar manner. We consider the correlation function $\left\langle \left(\phi_{i}^{*}\psi_{ja\alpha}\right)(p)\phi_{k}(p_{1})\overline{\psi}_{bl\dot{\alpha}}(-p_{2})\right\rangle$. The only diagram that is not a propagator correction is shown in figure \ref{fig:mixed_meson_diagram}. The rest of the computation is completely analogous.

The anomalous dimensions all turn out to be:
\begin{equation}
    \gamma_{\left( \phi^*\psi_\alpha\right)_{\text{sing}}} = \gamma_{\left( \phi^*\psi_\alpha\right)_{\text{adj}}} =
    \frac{1}{(4\pi)^2} \left[ 3e^2-\left( 2N_g+\frac{1}{2}\right)y^2\right].
\end{equation}
Evaluating this for $N_g=\frac{1}{2},\;\epsilon=\frac12$ and general $N_f$ in the standard $\epsilon$-expansion, we find
\begin{align}
    \gamma_{\left( \phi^*\phi\right)_{\text{adj}}}= &  \frac{510+263N_f +34N_f^2 \pm\sqrt{-1500-852N_{f}-71N_{f}^{2}+28N_{f}^{3}+4N_{f}^{4}}}{8N_{f}\left(4+N_{f}\right)\left(7+2N_{f}\right)}
    \\
    \gamma_{\left( \phi^*\phi\right)_{\text{sing}}}=  & \frac{1}{8N_{f}\left(4+N_{f}\right)\left(7+2N_{f}\right)}
    \left\{ 510+293N_f +49N_f^2+2N_f^3 \right. \nonumber \\ 
    & \left. \pm \left(1+N_{f}\right) \sqrt{-1500-852N_{f}-71N_{f}^{2}+28N_{f}^{3}+4N_{f}^{4}} \right\}, \\
    \gamma_{\left( \phi^*\psi_\alpha\right)_{\text{all}}} & = \frac{3\left(N_{f}+4\right)}{2N_{f}\left(2N_{f}+7\right)}.
\end{align}
These are plotted in figure \ref{fig:gamma_mesons}.
At both fixed points $\lambda_{0*}^\pm$, the anomalous dimensions of $\left( \phi^*\phi\right)_{\text{adj}}$ and $\left( \phi^*\psi_\alpha\right)$ tend to 0 as $N_f\rightarrow\infty$. However, $\gamma_{\left( \phi^*\phi\right)_{\text{sing}}}$ only approaches 0 at the $\lambda_{0*}^-$ fixed point, whereas in the $\lambda_{0*}^+$ fixed point it approaches the finite limit $\frac{1}{4}$.
The results at $\lambda_{0*}^-$ the fixed point agree with SUSY and with \cite{Benvenuti:2019}. By contrast, the limit of $\gamma_{\left( \phi^*\phi\right)_{\text{sing}}}$ at the $\lambda_{0*}^+$ fixed point deviates from them both.

We also show the anomalous dimensions of mesons in the revised $\epsilon$-expansion in figure \ref{fig:gamma_mesons_revised}. Here, only the $\lambda_{0*}^+$ gives a perturbatively stable fixed point, and $\lim_{N_f\rightarrow\infty}\gamma_{\left( \phi^*\phi\right)_{\text{sing}}}=\frac14$ as in the standard $\epsilon$-expansion. This is a violation of SUSY also in the large $N_f$ limit.

\end{appendix}

\bibliography{refs}
	\bibliographystyle{JHEP.bst}

\end{document}